\documentclass[%
 reprint,
 amsmath,amssymb,
 aps,
]{revtex4-1}
\usepackage{url}
\usepackage[colorlinks,linkcolor=blue]{hyperref}
\usepackage{graphicx}
\usepackage{dcolumn}
\usepackage{bm}

\begin{document}

\title{Robust and highly-efficient discrimination of chiral molecules\\
through three-mode parallel paths}
\author{Jin-Lei Wu$^{1}$}\author{Yan Wang$^{1}$}\author{Jie Song$^{1}$}\email[]{jsong@hit.edu.cn}\author{Yan Xia$^{2}$}\author{Shi-Lei Su$^{3}$}\author{Yong-Yuan Jiang$^{1}$}
\affiliation{$^{1}$School of Physics, Harbin Institute of Technology, Harbin 150001, China\\
$^{2}$Department of Physics, Fuzhou University, Fuzhou 350002, China\\
$^{3}$School of Physics, Zhengzhou University, Zhengzhou, 450001, China}

\begin{abstract}
We propose to discriminate chiral molecules by combining one- and two-photon processes in a closed-loop configuration. The one-photon-coupling intrinsic $\pi$-phase difference between two enantiomers leads to their different superposition states, which is then followed by a two-photon process through three-mode parallel paths~(3MPPs), enabling the discrimination of enantiomers by inducing their entirely-different population distributions. The 3MPPs are constructed by ``chosen paths", a method of shortcuts to adiabaticity~(STA), exhibiting a fast two-photon process. As an example, we propose to perform the scheme in $1,2$-propanediol molecules, which shows relatively robust and highly-efficient results under considering the experimental issues concerning unwanted transitions, imperfect initial state, pulse shaping, control errors and the effect of energy relaxations. The present work may provide help for laboratory researchers in a robust separation of chiral molecules.
\end{abstract}
\maketitle

\section{Introduction}
Chiral molecules~\cite{Pasteur1,Pasteur2}, involving two enantiomers~(a pair of a chiral molecule and its mirror image) that share most physical and chemical properties, own divergent functionality or activity for living matter dependent on the environment where they are present, so the discrimination and purification of (racemic) mixtures of enantiomers are of a strong necessity.
Chiral discrimination and purification are among the most difficult tasks in chemistry~\cite{Boden,McKendry,Rikken,Zepik,Bielski1,Bielski2}. Owing to the property of broken symmetry, one-photon and two-photon processes can coexist in a three-state configuration of chiral molecules, based on which physical~(especially optical) methods~\cite{Shapiro2000} have been becoming a promising alternative instead of chemical techniques, providing time-saving, convenient, and economical enantiomer-separation. Methods based on a closed-loop three-state ($\Delta$-type) system with microwave-driven rotational transitions is important and interesting~\cite{Kral2001,Kral2003,Li2007,Li2008,Li2010,Vitanov2019,Patterson2013,Patterson2013PRL,Patterson2014,Shubert2014,Lobsiger2015,Shubert2016,Eibenberger2017,Perez2017,Perez2018,Domingos2018}.
All above theoretical~\cite{Kral2001,Kral2003,Li2007,Li2008,Li2010,Vitanov2019} and experimental~\cite{Patterson2013,Patterson2013PRL,Patterson2014,Shubert2014,Lobsiger2015,Shubert2016,Eibenberger2017,Perez2017,Perez2018,Domingos2018} schemes base on the fact that the combined quantity defined by the triple product of three dipole-moment components is of opposite sign between enantiomers.

Theoretically, for example, Kr\'{a}l \emph{et al.}~\cite{Kral2001,Kral2003,Kral2007} proposed a method of coherently controlled adiabatic passage for achieving chiral separation, termed ``cyclic population transfer"~(CPT).
In CPT schemes, the interference of two-path stimulated Raman adiabatic passages~(STIRAP)~\cite{Bergmann1998,Vitanov2017} with zero~\cite{Kral2001} or nonzero ~\cite{Kral2003} detuning results in disparate population distributions between two enantiomers, depending on the total (intrinsic and optical) phase of the three coupling. In 2008, Li and Bruder~\cite{Li2008} proposed a fast nonadiabatic resonant pulse scheme. This is a three-stage scheme including sequential one-, two-, and one-photon processes, based on which left- and right-handed chiral molecules starting in a same initial state can evolve into different final states. Very recently, Vitanov and Drewsen~\cite{Vitanov2019} reported a scheme of the detection and separation of chiral molecules. This work applies the shortcuts to adiabaticity~(STA)~\cite{Bergmann1997,Berry2009,Chen2010,Wu2017} with a ``counterdiabatic field"~\cite{Demirplak2003} that plays a double-face role counteracting the nonadiabatic coupling for one enantiomer but strengthening the nonadiabatic coupling for another one, and therefore enables the $100\%$ enantiomer contrast in a particular state population. Experimentally, in samples of cold gas-phase molecules, Patterson \emph{et al.} first verified enantiomer differentiation by mapping the enantiomer-dependent sign of an electric dipole Rabi frequency onto the phase of free-induced decay signals in a dc field-assisted scheme~\cite{Patterson2013} and first demonstrated the technique of microwave three-wave mixing~(M3WM) for the sensitive chiral analysis including the probe of enantiomeric excess in a double-resonance scheme~\cite{Patterson2013PRL}. Then Shubert \emph{et al.}~\cite{Shubert2014}, combining M3WM and broadband microwave spectroscopy, determined not only the enantiomeric excess but also the absolute configurations~(species) in a supersonic jet experiment. Recently, Eibenberger \emph{et al.}~\cite{Eibenberger2017,Perez2017} realized a phase-dependent state-specific enantiomeric enrichment in cryogenic buffer gas by using four pulses, the first three for CPT and the last one for population detection. Whereafter, P\'{e}rez \emph{et al.}~\cite{Perez2017} reported a supersonic jet experiment of phase-dependent enantiomer-selective population enrichment in a novel microwave five-pulse scheme.

Fast, accurate, and robust manipulations always play a center role for all kinds of physical~(quantum) tasks, so analytical and numerical methods of sorts have been proposed to design control fields within recent decades, such as adiabatic passages~\cite{Bergmann1998,Vitanov2017}, composite pulse sequences~\cite{PhysRevA.83.053420,PhysRevLett.113.043001}, and quantum optimal control~\cite{PhysRevLett.104.083001,Glaser2015}.
As a control technique, STA provides another insight of pulse engineering, faster than adiabatic passages, requiring less pulses than composite pulse sequences, and holding analytical pulse shapes. Here we propose a scheme of discriminating chiral molecules by combining one- and two-photon processes in a closed-loop configuration. A $\pi/2$ pulse firstly used for the one-photon process and STA subsequently in the two-photon process enables three-mode parallel paths~(3MPPs) and induces entirely-different population distributions between enantiomers. The scheme is faster than the adiabatic CPT schemes~\cite{Kral2001,Kral2003}. Different from the fast nonadiabatic resonant pulse scheme~\cite{Li2008,Eibenberger2017,Perez2017}, the fast two-photon coherent population transfer is performed by speeding up a STIRAP with shaped pulses, not only fast but also robust against control errors. In contrast with the recent STA scheme~\cite{Vitanov2019} that performs simultaneously one- and two-photon processes with three pulses holding the strict match relation of time and amplitude, the present one carries out stepwise one- and two-photon processes, which loosens the match relation among the pulses and thus strengthens the robustness against control errors.

As an example, we propose to perform the scheme proposed here for the discrimination between the enantiomeric pair of a conformer of $1,2$-propanediol molecules, and consider the experimental issues including unwanted transitions, imperfect initial state, pulse shaping, control errors and the effect of energy relaxations. Relatively robust and highly-efficient results can be obtained.

The remainder of the paper is structured as follows. In Sec.~\ref{S2}, we give a STIRAP scheme of a population contrast between enantiomers and illustrate the adiabatic construction of 3MPPs. In Sec.~\ref{S3}, we propose a fast STA scheme for accelerating the STIRAP scheme by constructing alternative 3MPPs, containing analytical description and pulse engineering. In Sec.~\ref{S4}, the pulse overlap between the one- and two-photon processes and the phase sensitivity of the scheme are discussed. As an example, we propose to perform the scheme in $1,2$-propanediol molecules in Sec.~\ref{S5}. Finally, the conclusion appears in Sec.~\ref{S6}.

\section{Population contrast via STIRAP}\label{S2}
\begin{figure}\centering
\centering
\includegraphics[width=0.7\linewidth]{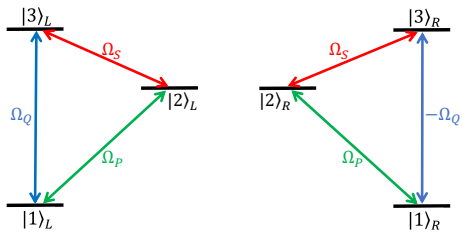}
\caption{Comparison of the coupling scheme between three discrete energy states in molecules with $L$~(on the left) and $R$~(on the right) handedness.}\label{f1}
\end{figure}
As shown in Fig.~\ref{f1}, the enantiomers, $L$- and $R$-handed molecules~($L$ and $R$ refer to left and right, respectively), are with a closed-loop configuration among three discrete energy states $|1\rangle$, $|2\rangle$ and $|3\rangle$. The electric-dipole-allowed $j$~($j=P,~S,~Q$) transition with dipole  moment $\overrightarrow{\mu_j}$ is driven by a field $\overrightarrow{E_j}= \overrightarrow{e_j}{\varepsilon}_j\cos(\omega_jt+\phi_j)$, where $\overrightarrow{e_j}$, ${\varepsilon}_j$, $\omega_j$, and $\phi_j$ are the unit vector, amplitude, frequency, and phase, respectively.
Then the Hamiltonian of such a closed-loop configuration can be represented in a matrix with basis vectors $\{|1\rangle,~|2\rangle,~|3\rangle\}$ (using the natural unit $\hbar=1$ and $|1\rangle$ being the zero-energy point)
\begin{eqnarray}\label{e1}
H_0=
\left(
\begin{array}{ccc}
0&\overrightarrow{\mu_P}\cdot\overrightarrow{E}_P &\overrightarrow{\mu_Q}\cdot\overrightarrow{E}_Q\\
\overrightarrow{\mu_P}\cdot\overrightarrow{E}_P&\omega_{1,2} &\overrightarrow{\mu_S}\cdot\overrightarrow{E}_S\\
\overrightarrow{\mu_Q}\cdot\overrightarrow{E}_Q&\overrightarrow{\mu_S}\cdot\overrightarrow{E}_S &\omega_{1,3}\\
\end{array}
\right),
\end{eqnarray}
in which $\omega_{1,n}~(n=2,3)$ is the $|1\rangle\leftrightarrow|n\rangle$ transition frequency.
The Rabi frequency of $j$ transition is defined as $\Omega_j=\overrightarrow{\mu_j}\cdot\overrightarrow{e_j}{\varepsilon}_j$, and we specify our model by choosing two enantiomers possessing identical $\Omega_P$ and $\Omega_S$ while opposite-sign $\Omega_Q$. Hereinafter the rotating-wave approximation condition $\omega_{j}\gg|\Omega_j|$ is taken into account.

\subsection{Fast one-photon process}
\begin{figure*}\centering
\centering
\includegraphics[width=0.7\linewidth]{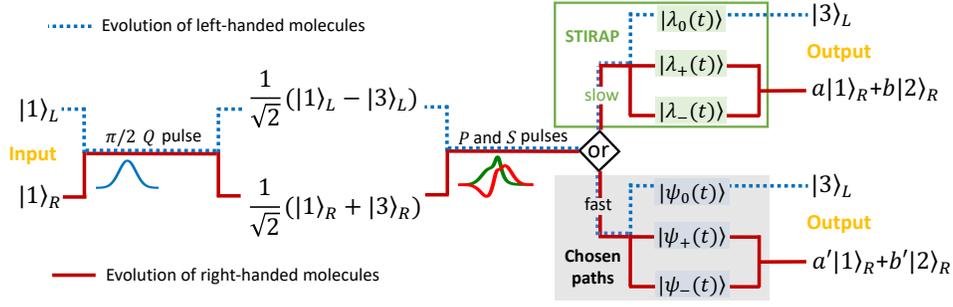}
\caption{Schematic diagram for the state evolutions of the enantiomers.}\label{f2}
\end{figure*}
The state evolutions of the enantiomers are illustrated in Fig.~\ref{f2}. The enantiomers are both prepared initially in $|1\rangle$. The scheme of the enantio-discrimination can be divided to two stages. In the first stage, only $Q$ pulse woks for the one-photon transition $|1\rangle\leftrightarrow|3\rangle$, which can be described by the Hamiltonian in the interaction picture
\begin{eqnarray}\label{e2}
H_1^{L,R}=\pm\frac12\Omega_Qe^{i\phi_q}|1\rangle\langle3|+{\rm H.c.},
\end{eqnarray}
for which we have considered the resonant condition $\omega_{1,3}=\omega_{Q}$.

After adopting a $\pi/2$ $Q$ pulse~(i.e., $\int\Omega_Qdt=\pi/2$) with $\phi_q=\pi/2$, the states of the $L$- and $R$-handed molecules are driven from $|1\rangle$ to, respectively,
$\frac{1}{\sqrt{2}}\left(|1\rangle_{L}-|3\rangle_{L}\right)$ and $\frac{1}{\sqrt{2}}\left(|1\rangle_{R}+|3\rangle_{R}\right)$. It is a fast non-adiabatic process and can be achieved by using a constant-amplitude rectangular pulse or a time-dependent shaped pulse. For example, we can choose a Gaussian pulse
\begin{eqnarray}\label{e3}
\Omega_Q=\Omega^q_{0} e^{-(t-t_{f1}/2)^{2} / T_q^{2}},
\end{eqnarray}
where $\Omega^q_{0}$ and $T_q$ are the maximum amplitude and width of the Gaussian pulse, respectively.
Here we set that the first stage starts at the time $t=0$ and ends at the time $t_{f1}=6T_q$ so as to cover the (almost) full pulse area $\int_0^{{t_{f1}}}\Omega_Qdt=\sqrt\pi\Omega^q_{0}T_q$, and $\Omega^q_{0}T_q=\sqrt\pi/2$.

\subsection{Two-photon process of STIRAP}
In the second stage, the interaction-picture Hamiltonian reads
\begin{eqnarray}\label{e4}
H=\frac12(\Omega_Pe^{i\phi_p}|1\rangle\langle2|+\Omega_Se^{i\phi_s}|2\rangle\langle3|)+{\rm H.c.},
\end{eqnarray}
where the resonant conditions $\omega_{1,2}=\omega_{P}$ and $\omega_{2,3}\equiv\omega_{1,3}-\omega_{1,2}=\omega_{S}$ are considered. The Hamiltonian~(\ref{e4}) is common for the $L$- and $R$-handed molecules, and its three instantaneous eigenstates with corresponding eigenenergies $\pm\Omega/2$ and 0, respectively, are
\begin{eqnarray}\label{e5}
\left|\lambda_{ \pm}(t)\right\rangle&=&\frac{1}{\sqrt{2}}\left[e^{i(\phi_p+\phi_s)}\sin \theta|1\rangle \pm e^{i\phi_s}|2\rangle+\cos \theta|3\rangle\right],\nonumber\\ \left|\lambda_{0}(t)\right\rangle&=&e^{i(\phi_p+\phi_s)}\cos \theta|1\rangle-\sin \theta|3\rangle,
\end{eqnarray}
with $\Omega\equiv\sqrt{\Omega_{P}^{2}+\Omega_{S}^{2}}$ and $\theta(t)\equiv\arctan \left[\Omega_{P}(t) / \Omega_{S}(t)\right]$.
For simplicity, we adopt $\phi_p=\phi_s=0$.
Under the adiabatic criterion $\dot{\theta}\ll\Omega$~\cite{Bergmann1998,Vitanov2017} which requires that $\Omega_P$ and $\Omega_S$ vary very slowly,  the nonadiabatic coupling among $|\lambda_{ \pm,0}(t)\rangle$ can be neglected. Then the couplings among \{$\left|\lambda_{\pm,0}\right\rangle$\} are avoided, and the 3MPPs are formed (as shown in the green box in Fig.~\ref{f2}).

In order to achieve the discrimination of chiral molecules with the states $|\Psi\rangle_L=\frac{1}{\sqrt{2}}\left(|1\rangle_{L}-|3\rangle_{L}\right)$ and $|\Psi\rangle_R=\frac{1}{\sqrt{2}}\left(|1\rangle_{R}+|3\rangle_{R}\right)$, we can follow the concept of the fractional STIRAP~\cite{Marte1991,Vitanov1999}. Assuming the second stage starts at the time $t_i$ and ends at the time $t_f$ and considering $\theta(t_i)=\pi/4$~(corresponding to $\Omega_P=\Omega_S$), the $L$-handed molecules will evolve along the dark state $|\lambda_{0}(t)\rangle$, while the $R$-handed molecules along the equal-weighted superposition of  $\exp(\frac{i}2\int\Omega dt)|\lambda_{+}(t)\rangle$ and $\exp(-\frac{i}2\int\Omega dt)|\lambda_{-}(t)\rangle$~(here the geometric phase~\cite{Berry1984} is zero). Then we choose $\theta(t_f)=\pi/2$~(corresponding to $\Omega_P\gg\Omega_S$), and the final states of the $L$- and $R$-handed molecules, respectively, are
\begin{eqnarray}\label{e6}
|\Psi\rangle_L&=&-|3\rangle_{L},\nonumber\\
|\Psi\rangle_R&=&\cos\mathcal{A}|1\rangle_R+i\sin\mathcal{A}|2\rangle_R,
\end{eqnarray}
with $\mathcal{A}\equiv\frac12\int_{t_i}^{t_f}\Omega dt$. That is, $L$- and $R$-handed molecules finally hold entirely different population distributions of $|3\rangle$, $P_{3L}\equiv|\langle3|\Psi\rangle_L|^2=1$ but $P_{3R}\equiv|\langle3|\Psi\rangle_R|^2=0$.

The shapes of $P$ and $S$ pulses used in the STIRAP process can be chosen without limitation, as long as the boundary conditions $\Omega_P(t_i)=\Omega_S(t_i)$ and $\Omega_P(t_f)\gg\Omega_S(t_f)$ are satisfied within the adiabatic criterion. Here we choose, for example, a double-Gaussian $P$ pulse and a single-Gaussian $S$ pulse, respectively
\begin{eqnarray}\label{e7}
\Omega_P&=&\Omega_{0} e^{-[(t-t_{i})-(t_{f}-t_{i}-\tau)/2]^{2} / T^{2}}\nonumber\\&&+\Omega_{0} e^{-[(t-t_{i})-(t_{f}-t_{i}+\tau)/2]^{2} / T^{2}},\nonumber\\
\Omega_S&=&\Omega_{0} e^{-[(t-t_{i})-(t_{f}-t_{i}-\tau)/2]^{2} / T^{2}},
\end{eqnarray}
where $\Omega_{0}$ and $T$ are the maximum amplitude and width of a single Gaussian pulse, respectively. $\tau$ is the delay between the two single Gaussian pulses of $\Omega_P$. Here $t_f=t_i+6T+\tau$ can be set so as to guarantee that all pulse areas are almost covered.

As is well known, however, the STIRAP-based scheme is pretty slow because of the limitation of the adiabatic criterion. Such a slow enantio-discrimination is of low efficiency, and may be ineffective due to the relaxations of higher-energy states and the accumulation of control errors.

\section{Enantio-discrimination via chosen paths}\label{S3}
\subsection{Analytical description}
The slow enantio-discrimination in the STIRAP scheme can be speeded up by ``chosen paths"~(CP), a method of STA, proposed by the first author and his co-workers~\cite{Wu2017}. The CP scheme, choosing three appropriate dressed states as 3MPPs instead of $|\lambda_{\pm,0}(t)\rangle$, has two key points: (i) The evolution in the second stage bases on the two-photon process $|1\rangle\leftrightarrow|2\rangle\leftrightarrow|3\rangle$ in which each one-photon transition is resonant; (ii) There is no coupling among the chosen paths.

We use the Hamiltonian that satisfies the point (i), as
\begin{eqnarray}\label{e8}
H_c=\frac12(\Omega^c_P|1\rangle\langle2|+\Omega^c_S|2\rangle\langle3|)+{\rm H.c.},
\end{eqnarray}
with the modified Rabi frequencies $\Omega^c_P=\Omega_P+\Omega_1$ and $\Omega^c_S=\Omega_S+\Omega_2$, where $\Omega_{P,S}$ are the STIRAP pulses in Eq.~(\ref{e4}), and $\Omega_{1,2}$ can be considered the counterdiabatic fields. The 3MPPs satisfying the orthogonality and completeness can be chosen as
\begin{eqnarray}\label{e9}
|\psi_{0}(t)\rangle&=&\cos\beta\cos\theta|1\rangle-i\sin\beta|2\rangle-\cos\beta\sin\theta|3\rangle,\nonumber\\
|\psi_{\pm}(t)\rangle&=&\frac{1}{\sqrt{2}}[(\sin \theta \mp i \sin \beta \cos \theta)|1\rangle \pm \cos \beta(t)|2\rangle\nonumber\\
&&+(\cos \theta \pm i \sin \beta \sin \theta)|3\rangle].
\end{eqnarray}
With $|\psi_{\pm,0}(t)\rangle$ being the evolution paths, the conditions of the enantio-discrimination are
\begin{eqnarray}\label{e10}
\theta(t_i)=\pi/4,~\theta(t_f)=\pi/2,~\beta(t_i)=\beta(t_f)=0,
\end{eqnarray}
in which the last one ensures the coincidence between $|\psi_{\pm,0}(t)\rangle$ and $|\lambda_{\pm,0}(t)\rangle$ at the time $t_i$ and $t_f$.
It is convenient to move $H_c$ to the frame with the time-independent chosen paths being basis by the unitary operator $U_{0}=\sum_{n=\pm, 0}|\psi_{n}\rangle\langle\psi_{n}(t)|$~\cite{Baksic2016}, and thus $H_c$ becomes
\begin{eqnarray}\label{e11}
H'_c&=&U_{0} H U_{0}^{\dagger}-i U_{0} \dot{U}_{0}^{\dagger}\nonumber\\
&=&\frac12[\xi(|\psi_{+}\rangle\langle\psi_{+}|-|\psi_{-}\rangle\langle\psi_{-}|)\nonumber\\
&&+(\xi_+|\psi_{0}\rangle\langle\psi_{+}|+\xi_-|\psi_{0}\rangle\langle\psi_{-}|+{\rm H.c.})/\sqrt2],
\end{eqnarray}
with
\begin{eqnarray}\label{e12}
\xi&=& \cos\beta(\Omega_1\sin\theta+\Omega_2\cos\theta+\Omega)+\dot{\theta}\sin\beta\nonumber\\
\xi_\pm&=&i[\sin\beta(\Omega_1\sin\theta+\Omega_2\cos\theta+\Omega)-2\dot\theta\cos\beta]\nonumber\\
&&\pm(\Omega_1\cos\theta-\Omega_2\sin\theta-2\dot{\beta}).
\end{eqnarray}
According to the point (ii), $\xi_\pm$ must be zero, which can be solved by the modified Rabi frequencies with the expressions
\begin{eqnarray}\label{e13}
\Omega^c_P&=&2(\dot{\beta}\cos\theta+\dot\theta\cot\beta\sin\theta),\nonumber\\
\Omega^c_S&=&-2(\dot{\beta}\sin\theta-\dot\theta\cot\beta\cos\theta).
\end{eqnarray}
By using the Rabi frequencies in Eq.~(\ref{e13}), the evolution will follow the 3MPPs $|\psi_{\pm,0}(t)\rangle$~(as shown in the shadow in Fig.~\ref{f2}).
With the states $|\Psi\rangle_L=\frac{1}{\sqrt{2}}\left(|1\rangle_{L}-|3\rangle_{L}\right)$ and $|\Psi\rangle_R=\frac{1}{\sqrt{2}}\left(|1\rangle_{R}+|3\rangle_{R}\right)$ and the conditions in Eq.~(\ref{e10}), through 3MPPs $|\psi_{\pm,0}(t)\rangle$ the final states of the $L$- and $R$-handed molecules, respectively, are
\begin{eqnarray}\label{e14}
|\Psi\rangle_L&=&-|3\rangle_{L},\nonumber\\
|\Psi\rangle_R&=&\cos\mathcal{A}^{c}|1\rangle_R+i\sin\mathcal{A}^{c}|2\rangle_R,
\end{eqnarray}
where we define $\mathcal{A}^{c}\equiv\frac12\int_{t_i}^{t_f}\xi dt$. Equation~(\ref{e14}) has the same form as Eq.~(\ref{e6}), indicating that $L$- and $R$-handed molecules finally possess entirely different population distributions of $|3\rangle$. Different from the STIRAP scheme, the enantio-discrimination in the CP scheme can be fast, not limited by the adiabatic criterion.

\subsection{Pulse engineering}
\begin{figure}[htb]\centering
\includegraphics[width=0.9\linewidth]{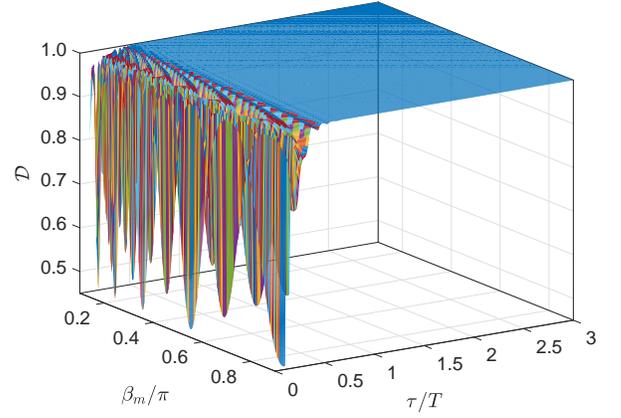}
\caption{Dependence of $\mathcal{D}$ on $\tau/T$ and $\beta_m$. Parameters: $t_f=t_i+6T+\tau$ and $T_\beta=(t_f-t_i)/6$.}\label{f3}
\end{figure}
There are diverse pulse schemes of satisfying the conditions in Eq.~(\ref{e10}). For instance, based on the definition of $\theta$ in Eq.~(\ref{e5}) and the STIRAP pulses in Eq.~(\ref{e7}), $\theta(t_i)=\pi/4$ and $\theta(t_f)=\pi/2$ can be obtained. $\beta(t_i)=\beta(t_f)=0$ can be ensured with a Gaussian function
\begin{eqnarray}\label{e15}
\beta=\beta_me^{-[(t-t_i)-(t_f-t_i)/2]^2/T^2_\beta},
\end{eqnarray}
where $T_\beta=(t_f-t_i)/6$ is set. Therewith, the enantio-discrimination of the second stage in the CP scheme will depend on the values of $\tau/T$ and $\beta_m$.

We define a quantity,
\begin{eqnarray}\label{e16}
\mathcal{D}=\left|P_{3L}-P_{3R}\right|
\end{eqnarray}
to measure the discrimination of chiral molecules with $\mathcal{D}\in[0,1]$,
serving as the fidelity of performing discrimination of chiral molecules. With the second-stage initial states $|\Psi\rangle_L=\frac{1}{\sqrt{2}}\left(|1\rangle_{L}-|3\rangle_{L}\right)$ and $|\Psi\rangle_R=\frac{1}{\sqrt{2}}\left(|1\rangle_{R}+|3\rangle_{R}\right)$, the dependence of $\mathcal{D}$ on $\tau/T$ and $\beta_m$ can be illustrated, as Fig.~\ref{f3}, by means of solving numerically the Schr\"{o}dinger equation with respect to the CP-scheme Hamiltonian Eq.~(\ref{e8}). As shown clearly in Fig.~\ref{f3}, $\mathcal{D}$ will keep unchanged with varying $\tau/T$ and $\beta_m$ and maintain in unity if the ratio $\tau/T$ is over about $0.5$ within $\beta_m\in(0,\pi)$, which means that the scheme possesses great flexibility in parameter selections. The parameter $\tau$ that affects the initial value of $\Omega_P$ can not be too small, because a small delay between the two single Gaussian pulses of $\Omega_P$ will lead to $\Omega_P(t_i)>\Omega_S(t_i)$ and then deviate the condition $\theta(t_i)=\pi/4$.

\begin{figure*}\centering
\includegraphics[width=0.6\linewidth]{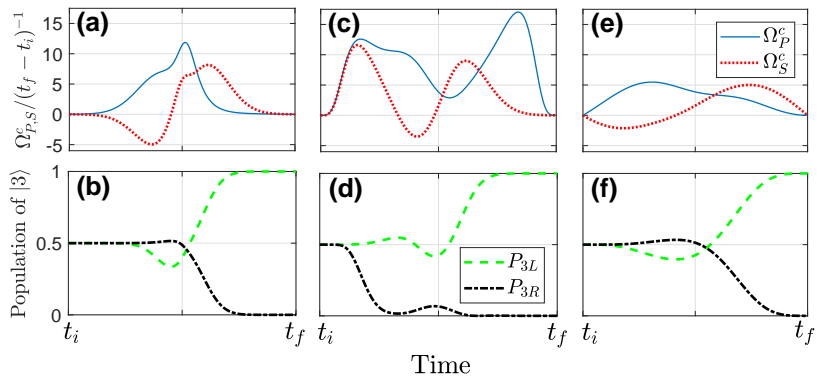}
\caption{(a), (c), and (e): Shapes of $\Omega^c_P$ and $\Omega^c_S$; (b), (d), and (f): Second-stage population evolutions of $|3\rangle$. Parameters for (a) and (b): $\tau=2T$, $\beta_m=0.25\pi$, $t_f=t_i+6T+\tau$, and $T_\beta=(t_f-t_i)/6$ in Eqs.~(\ref{e7}) and (\ref{e15}); Parameters for (c) and (d): $\beta_m=0.25\pi$ and $T_\beta=(t_f-t_i)/6$ in Eq.~(\ref{e15}); Parameters for (e) and (f): $\beta_m=0.25\pi$ and $\epsilon=0.001$ in Eq.~(\ref{e18}).}\label{f4}
\end{figure*}
With $\tau=2T$ and $\beta_m=0.25\pi$, the shapes of $\Omega^c_P$ and $\Omega^c_S$ are shown in Fig.~\ref{f4}(a), based on which then the second-stage evolutions of enantiomers can be obtained, and the population evolutions of $|3\rangle$ are plotted in Fig.~\ref{f4}(b) showing the full enantio-discrimination. Figure~\ref{f4}(a) just shows an example of pulse shapes and durations, while in fact pulse shapes and durations have great versatility of selection according to different values of $\tau/T$ and $\beta_m$. Furthermore, the form of $\theta$ can also be designed directly, instead of originating from the STIRAP pulses. For instance, $\theta$ can be chosen as
\begin{eqnarray}\label{e17}
\theta&=&\frac\pi4+\frac12\left[\frac{\pi(t-t_i)}{2(t_f-t_i)}-\frac13\sin\frac{2\pi(t-t_i)}{t_f-t_i}\right.\nonumber\\
&&\left.+\frac1{24}\sin\frac{4\pi(t-t_i)}{t_f-t_i}\right],
\end{eqnarray}
based on which and using $\beta$ in Eq.~(\ref{e15}) the shapes of $\Omega^c_P$ and $\Omega^c_S$ and the population evolutions of $|3\rangle$ can be depicted and shown in Figs.~\ref{f4}(c) and (d), respectively. Alternatively, $\beta$ in Eq.~(\ref{e15}) can be replaced by another function, for example a single-period $\cos$-like function as
\begin{eqnarray}\label{e18}
\beta=\left\{
\begin{array}{cc}
\frac{\beta_m}{2}\left[1-\cos\frac{2 \pi(t-t_i)}{t_f-t_i}\right]+\epsilon, &t\in[t_i,t_f]\\
0,&\text{otherwise}
\end{array}\right.,
\end{eqnarray}
where a small-value $\epsilon$~(e.g., $\epsilon=0.001$) is introduced to avoid infinite $\Omega^c_{P,S}$. Using $\theta$ in Eq.~(\ref{e17}) and $\beta$ in Eq.~(\ref{e18}), the shapes of $\Omega^c_P$ and $\Omega^c_S$ and the population evolutions of $|3\rangle$ are shown in Figs.~\ref{f4}(e) and (f), and the enantio-discrimination is still fully obtained.

The pulses designed in Figs.~\ref{f4}(a), (c), and (e) are smooth without any singularity and turned on~(off) at zero, all of which can be generated with an arbitrary waveform generator in experiment. The absolute values and signs of the pulses can be controlled by modulating amplitudes ${\varepsilon}_j$ and phases $\phi_j$ of the corresponding fields, respectively.
Beyond the pulse forms above, one can find abundant alternative forms eligible for the CP scheme by means of many pulse engineering scenarios~\cite{PhysRevA.89.043408,PhysRevA.92.062136,Kang2016SR,PhysRevA.97.033407,Wu_2019} or optimal control techniques~\cite{PhysRevLett.104.083001,Glaser2015,PhysRevA.85.023611,PhysRevLett.111.050404,PhysRevA.88.043422,PhysRevA.95.063403,PhysRevA.98.043421}.

\section{Pulse overlap and phase sensitivity}\label{S4}
On one hand, the present enantio-discrimination scheme bases on the combination of one- and two-photon processes. Such a discrimination of chiral molecules can make the one-photon process~($Q$ pulse) free from the real-time interference with the two-photon process~($P$ and $S$ pulses), which therefore holds greater robustness against pulse match errors in contrast with the simultaneous-three-pulse scheme~\cite{Vitanov2019}. On the other hand, phase sensitivity is a common property for the existing schemes of optical chiral discrimination~\cite{Kral2001,Kral2003,Li2007,Li2008,Li2010,Patterson2013,Patterson2013PRL,Patterson2014,Shubert2014,Lobsiger2015,Shubert2016,Eibenberger2017,Perez2017,Perez2018,Domingos2018,Vitanov2019}. The present one is also a phase-sensitive scheme, for which we use a $Q$ pulse with $\pi/2$ phase and $P$ and $Q$ pulses with zero phase~(not including the sign of the amplitude). In this section we investigate the pulse overlap between the two stages and the phase sensitivity of the present scheme.
\subsection{Pulse overlap between two stages}
In an ideal case, $P$ and $S$ pulses are supposed to be executed after the ending of $Q$ pulse, i.e., $t_i\leq t_{f1}$. In fact, however, a moderate pulse overlap between the two stage is acceptable for the full discrimination of chiral molecules owing to the using of the shaped pulses, and can besides shorten the operation time to some extent. Since the Gaussian $Q$ pulse decreases to~($P$ and $S$ pulses increase from), gradually, a zero amplitude with a zero slope, there are two periods, one before $t_{f1}$ when the desired one-photon process has been completed and the other when the two-photon process has almost not started after $t_{i}$. Therefore, a pulse overlap between the two stages covering such two periods has little effect on the entirely-different population distributions of two enantiomers, which is different from the stepwise schemes~\cite{Li2008,Eibenberger2017,Perez2017} that are sensitive to the interstage pulse overlap .

\begin{figure*}\centering
\includegraphics[width=0.75\linewidth]{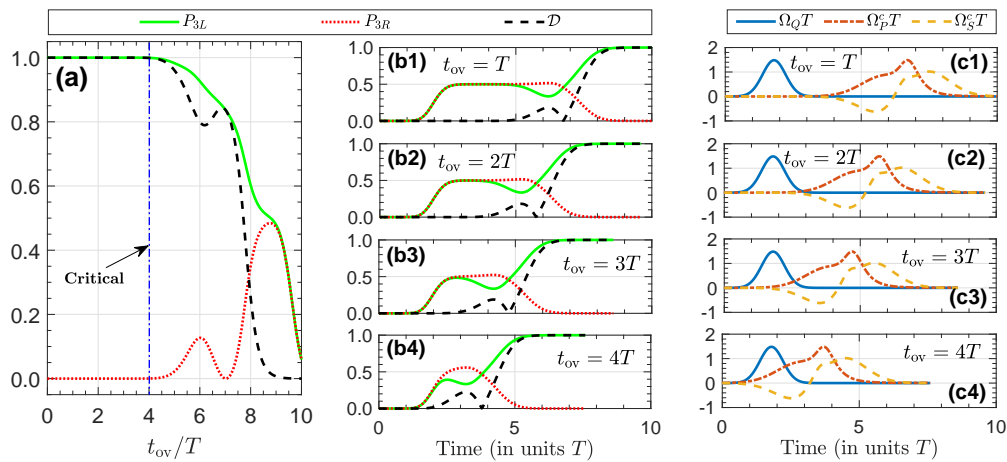}
\caption{(a)~Effect of the interstage overlap time $t_{\rm ov}$ on the enantio-discrimination. (b1)-(b4)~Population evolutions of $|3\rangle$ of two enantiomers with different $t_{\rm ov}$. (c1)-(c4)~Shapes of the three Rabi frequencies with different $t_{\rm ov}$. Parameters: $\tau=2T$, $\beta_m=0.25\pi$, $t_f=t_i+6T+\tau$, $T_\beta=(t_f-t_i)/6$, $\Omega_0^q=1.481/T$, $T_q=0.5983T$ and $t_{f1}=3.590T$.}\label{f5}
\end{figure*}
If $t_i$ is much earlier than $t_{f1}$ or even $t_i\leq0$ which induces the real-time interference between the one- and two-photon processes, will the full enantio-discrimination be achieved finally? In the following, we investigate the acceptable maximum overlap time, using $Q$ pulse in Eq.~(\ref{e3}) and $P$ and $S$ pulses shown in Fig.~\ref{f4}(a). The maximum amplitude of $P$ and $S$ pulses is $\max\{\Omega^c_P,~\Omega^c_S\}=11.85/(t_f-t_i)=1.481/T$, and we suppose that the two stages hold the equal maximum amplitude $\Omega_0^q=1.481/T$ that gives $T_q=0.5983T$ and then $t_{f1}=3.590T$. The interstage overlap time is defined as $t_{\rm ov}=t_{f1}-t_i$, and then the effect of $t_{\rm ov}$ on the enantio-discrimination can be depicted by means of the Schr\"{o}dinger equation with respect to $H_1^{L,R}$ in Eq.~(\ref{e2}) plus $H$ in Eq.~(\ref{e4}) and the initial state $|1\rangle$, as shown in Fig.~\ref{f5}(a). Figure~\ref{f5}(a) shows an apparent critical value around $t_{\rm ov}=4T$ that divides the full enantio-discrimination into an valid region~($t_{\rm ov}\leq4T$) and an invalid region~($t_{\rm ov}>4T$). Then we pick up different $t_{\rm ov}$ in the valid region and plot the population evolutions of $|3\rangle$ of two enantiomers in Figs.~\ref{f5}(b1)-(b4) and the corresponding shapes of the three Rabi frequencies in Figs.~\ref{f5}(c1)-(c4).

We can learn that for $t_{\rm ov}=T$ or $2T$, there is little pulse area overlap between the two stages, so no real-time interference between the one- and two-photon processes occurs during the enantio-discrimination. Nevertheless for $t_{\rm ov}\geq3T$, there is a finite pulse area overlap between the two stages and thus a little bit real-time interference between the one- and two-photon processes occurs. Even for $t_{\rm ov}> t_{f1}=3.590T$ that means that $P$ and $S$ pules are applied earlier than $Q$ pulse, the full enantio-discrimination can be obtained as shown in Fig.~\ref{f5}(b4). For $t_{\rm ov}\geq3.590T$, the scheme is in fact a simultaneous-three-pulse scheme but is still more robust against the pulse delay errors than the simultaneous-three-pulse scheme~\cite{Vitanov2019} and the stepwise schemes~\cite{Li2008,Eibenberger2017,Perez2017} because there exists a wide valid range from $t_{\rm ov}=3.590T$ to the critical value. In a word, the present scheme is very robust against the pulse delay error, and can still perform the full enantio-discrimination even if there is the real-time interference between the one- and two-photon processes, which mainly because that the interstage pulse area overlap is so insignificant that the real-time interference is very slight and hardly destroy the individual functions of the one and two-photon processes.

\subsection{Phase sensitivity}
\begin{figure}\centering
\includegraphics[width=\linewidth]{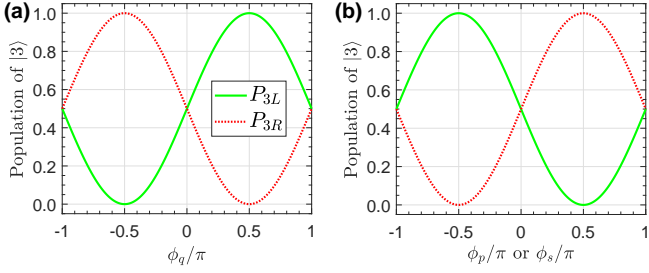}
\caption{Phase-dependent final population of $|3\rangle$ of two enantiomers with one pulse phase varying but other two being zero. Parameters are the same as Fig.~\ref{f5}(c2).}\label{f6}
\end{figure}

The enantio-discrimination in the present scheme depends on the intrinsic and optical phase of the three coupling in the closed-loop figuration, and the discussion above is established on the parameter setting for the phases of three pulses as $\phi_q=\pi/2$ and $\phi_p=\phi_s=0$ so far. As a matter of fact, the parameter setting for these phases is not monotonous. If all these phases are not assigned, the states of two enantiomers at the ending of the first stage are \begin{eqnarray}\label{e19}
|\Psi\rangle_{L,R}=\frac{1}{\sqrt{2}}\left[e^{i(\phi_q\mp\frac{\pi}2)}|1\rangle_{L}-|3\rangle_{L}\right].
\end{eqnarray}
Then through 3MPPs in the STIRAP scheme, the enantio-discrimination requires that one of two enantiomers evolves in the second stage along the dark state $|\lambda_{0}(t)\rangle$ but the other along the equal-weighted superposition of $\exp(\frac{i}2\int\Omega dt)|\lambda_{+}(t)\rangle$ and $\exp(-\frac{i}2\int\Omega dt)|\lambda_{-}(t)\rangle$. According to the forms of $|\lambda_{0,\pm}(t)\rangle$ in Eq.~(\ref{e5}), the condition of this requirement is \begin{eqnarray}\label{e20}
\phi_p+\phi_s-\phi_q=\left(n+\frac12\right)\pi,
\end{eqnarray}
with an integer number $n$, which shows a general three-pulse phase relation for a full enantio-discrimination. Likewise, this relation is also applicative in the CP scheme. To this end, by using the pulse shapes plotted in Fig.~\ref{f5}(c2), Fig.~\ref{f6} shows the phase sensitivity of the enantio-discrimination through considering the phase-dependent final population of $|3\rangle$ of two enantiomers. Figure~\ref{f6} demonstrates that one can make the entirely-different population distributions between $L$- and $R$-handed molecules by setting any two pulses with zero phase but the third one with a relative phase of $\pm\pi/2$, and the relative phase sign determines the excited enantiomer.

\section{Experimental consideration}\label{S5}
The present enantio-discrimination scheme can be applied to diverse samples of chiral molecules, and several species have been adopted in recent enantio-discrimination experiments, such as $1,2$-propanediol~\cite{Patterson2013,Patterson2013PRL,Eibenberger2017}, carvone~\cite{Shubert2014,Perez2017}, solketal~\cite{Lobsiger2015}, 4-carvomenthenol~\cite{Shubert2015}, menthone~\cite{Shubert2016,Perez2017}, and cyclohexylmethanol~\cite{Perez2018}, etc.

\subsection{Molecule candidate and master equation}
\begin{figure}\centering
\includegraphics[width=0.9\linewidth]{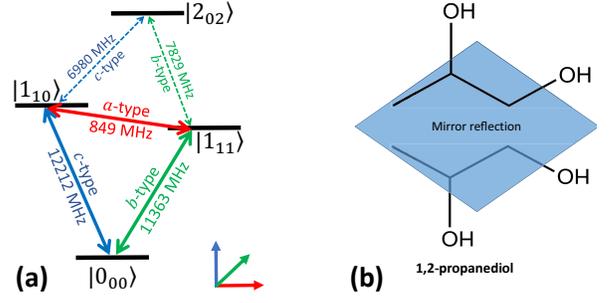}
\caption{(a)~Closed-loop configuration in an enantiomeric pair of a conformer of $1,2$-propanediol molecules~(solid thick double-headed arrows) and unwanted transitions~(dashed thin double-headed arrows). (b)~Mirror-refection molecular structure diagram of two enantiomers of $1,2$-propanediol.}\label{f7}
\end{figure}

We here take $1,2$-propanediol as an example to verify the scheme and discuss experimental issues. A closed-loop configuration~[$|0_{00}\rangle\leftrightarrow|1_{11}\rangle\leftrightarrow|1_{10}\rangle\leftrightarrow|0_{00}\rangle$, as shown in Fig.~\ref{f7}(a)] within $0.8-13$ GHz microwave regime comprises three rotational energy states of an enantiomeric pair of a conformer of $1,2$-propanediol molecules whose microwave spectrum can be found in Ref.~\cite{LOVAS200982}, which was already used in experiment~\cite{Patterson2013}. The mirror-refection molecular structure diagram of two enantiomers is plotted in Fig.~\ref{f7}(b).
The energy levels designated with $|J_{K_a,K_c}\rangle$, where $J$ is the rotational quantum number and $K_a$ and $K_c$ are the projections of $J$ onto the principal axes of the molecule. The adopted enantiomers of $1,2$-propanediol are of the conformer constants $A=5872.06$~MHz, $B=3640.11$~MHz, and $C=2790.97$~MHz, and have three types of rotational transitions, $a$-type, $b$-type, and $c$-type with dipole moments $\mu_a=1.201$~Debye, $\mu_b=1.916$~Debye, and $\mu_c=0.365$~Debye, respectively.

The coincidence between Figs.~\ref{f1}(a) and \ref{f7}(a) can be made by defining the states $|1\rangle\equiv|0_{00}\rangle$, $|2\rangle\equiv|1_{11}\rangle$ and $|3\rangle\equiv|1_{10}\rangle$, and the transition frequencies $\omega_{1,2}=11363$~MHz, $\omega_{1,3}=12212$~MHz and $\omega_{2,3}=849$~MHz. The $a$-type, $b$-type, and $c$-type transitions are driven resonantly by orthogonal $S$ pulse at frequency $\omega_S=849$~MHz, $P$ pulse at frequency $\omega_P=11363$~MHz, and $Q$ pulses at frequency $\omega_Q=12212$~MHz, respectively. We can choose appropriate axes such that
$\overrightarrow{\mu_j}$ and $\overrightarrow{E}_P$ have the same orientation, and two enantiomers have identical $\Omega_P$ and $\Omega_S$ but opposite-sign $\Omega_Q$. For such a specific experimental model, there may be unwanted couplings among possible~(off-resonant) rotational transitions and driving fields, and we consider the most likely two unwanted transitions $|1_{10}\rangle\leftrightarrow|2_{02}\rangle$ of frequency $6980$~MHz~($c$-type) and  $|1_{11}\rangle\leftrightarrow|2_{02}\rangle$ of frequency $7829$~MHz~($b$-type). Then the evolutions of the two enantiomers are dominated by the Hamiltonian
\begin{eqnarray}\label{e21}
H_{L,R}&=&H_0+\omega_4|4\rangle\langle4|+[\Omega_P\cos(\omega_Pt+\phi_p)|2\rangle\langle4|\nonumber\\
&&\pm \Omega_Q\cos(\omega_Qt+\phi_q)|3\rangle\langle4|+{\rm H.c.}],
\end{eqnarray}
where $H_0$ is given in Eq.~(\ref{e1}), $|4\rangle\equiv|2_{02}\rangle$ and $\omega_4=19192$~MHz. For the Rabi frequencies of megahertz order, the probabilities of the two unwanted off-resonant transitions are negligible indeed, because the detuning is so large that even over $3.5$~GHz.

There are usually two kinds of valid tools for cooling the molecular samples, buffer gas cooling and supersonic expansions. According to recent experiments, the molecular samples can be cooled to a temperature of around $5$-$10$~K by using a cryogenic buffer gas cell~\cite{Patterson2013,Patterson2013PRL,Eibenberger2017}, and the supersonic expansion can cool the molecules to rotational temperatures of about $1$-$2$~K~\cite{Shubert2014,Shubert2015,Perez2017,Perez2018}. In order to obtain the molecular sample of $1,2$-propanediol prepared initially in $|1\rangle$, the sample should be cooled as sufficiently as possible. Here we assume that the scheme is performed at relatively low temperature and all molecules in the sample of $1,2$-propanediol are prepared initially in a lowly-mixed state of the pure states $|1\rangle$, $|2\rangle$ and $|3\rangle$, with the density operator $\rho_0=0.998|1\rangle\langle1|+0.001|2\rangle\langle2|+0.001|3\rangle\langle3|$ that means $|2\rangle$ and $|3\rangle$ are mixed into the desired initial state $|1\rangle$ with the same probability $0.001$. Then the density operator $\rho$ at arbitrary time obeys the Markovian master equation
\begin{eqnarray}\label{e22}
\frac{\partial\rho}{\partial t}&=&-i[H_{L,R},~\rho]\nonumber\\
&&-\sum_{j=2,3}\frac{\gamma_{1,j}}{2}\left(\sigma_{1,j}^+\sigma_{1,j}^-\rho-2\sigma_{1,j}^-\rho\sigma_{1,j}^++\rho\sigma_{1,j}^+\sigma_{1,j}^-\right)
\nonumber\\
&&-\frac{\gamma_{2,3}}{2}\left(\sigma_{2,3}^+\sigma_{2,3}^-\rho-2\sigma_{2,3}^-\rho\sigma_{2,3}^++\rho\sigma_{2,3}^+\sigma_{2,3}^-\right),
\end{eqnarray}
where $\gamma_{m,n}$~($m,n=1,2,3;~m<n$) is the relaxation rate from $|n\rangle$ to $|m\rangle$, and $(\sigma_{m,n}^+)^\dag=\sigma_{m,n}^-\equiv|m\rangle\langle n|$.

\subsection{Pulse shaping}
\begin{figure}[htb]\centering
\includegraphics[width=0.9\linewidth]{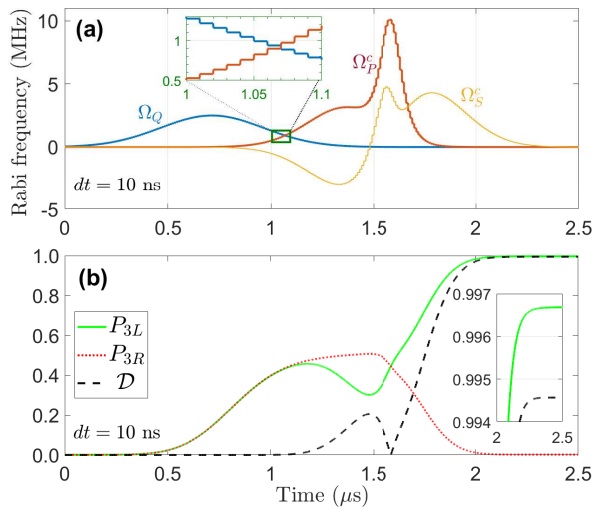}
\caption{(a)~Waveforms of the three pulses with a time resolution $dt=10$~ns. The inner plot denotes local zoom. (b)~Population evolutions of $|3\rangle$ of two enantiomers by using the pulse waveforms in (a). Parameters: $\Omega_0^q=2.5$~MHz, $T_q=0.3545~\mu$s, $t_{f1}=1.418~\mu$s, $T=0.15~\mu$s, $\tau=0.30~\mu$s, $t_i=0.618~\mu$s, $t_f=2.5~\mu$s, and $\gamma_{1,2}=\gamma_{1,3}=\gamma_{2,3}=0$.}\label{f8}
\end{figure}
The amplitudes of the Rabi frequencies can be controlled by modulating the corresponding filed amplitudes~(voltages applied to the
electro-optic modulators), and the sign change of $S$ pulse can be implemented by performing an instantaneous $\pi$-phase flip. The required respective phases of the three pulses can be determined by setting the phases of the corresponding fields. Recently, many microwave-regime experiments of the pulse shaping including the modulations of amplitudes, frequencies and phases~(flip) by means of arbitrary waveform generators have been reported~\cite{PhysRevLett.110.240501,BBZhou,PhysRevA.95.042345,Wang_2018,Zhang_2018,Felix,PhysRevApplied.11.034030}. The electric dipole moments of the considered transitions are of 1 Debye order, and it is securable to control the Rabi frequencies within $0$-$10$ MHz by using the maximum field strength around $\sim2~V/$cm. Here we pick up $\max\{\Omega_P^c,\Omega_S^c\}=10$~MHz and $\Omega_0^Q=2.5$~MHz to conduct the present enantio-discrimination scheme.

The used waveforms of the three pulses are varied continuously, which requires the infinite time resolution in theory. Although it is unrealistic to obtain an infinite time resolution in practice, a relatively long time resolution is still applicative for the high-$\mathcal{D}$ enantio-discrimination. For example in Fig.~\ref{f8}(a), we use a time resolution $dt=10$~ns, and suppose that each waveform consists of a series of ($10$~ns-duration) rectangular pulses. Based on these rectangular pulse sequences and the master equation~(\ref{e22}) with the initial (mixed) state $\rho_0$, the population evolutions of $|3\rangle$ of two enantiomers are plotted in Fig.~\ref{f8}(b) that shows a nearly full enantio-discrimination with $\mathcal{D}=0.9946$~(the energy relaxations of $|2\rangle$ and $|3\rangle$ are not considered for the moment).
The recent microwave-based experiments reported that arbitrary waveform generators can provide the minimal possible time resolution $\sim0.25$~ns~\cite{PhysRevLett.110.240501} and even $\sim0.1$~ns~\cite{BBZhou}. In the following, we set safely the time resolution of the three pulses as $dt=1$~ns.

\subsection{Control errors}
Perfect control of experimental operations is desired for a full enantio-discrimination but almost impossible, so it is essential to investigate the effect of control errors in the three pulses on the execution of the enantio-discrimination task. In this subsection, we mainly pay attention to two error sources: (1)~Systematic errors, including frequency drifts and amplitude drifts; (2)~Random amplitude noises, including additive white Gaussian noises~(AWGN) and random fluctuations.
\subsubsection{Systematic errors}
\begin{figure*}[htb]\centering
\includegraphics[width=0.7\linewidth]{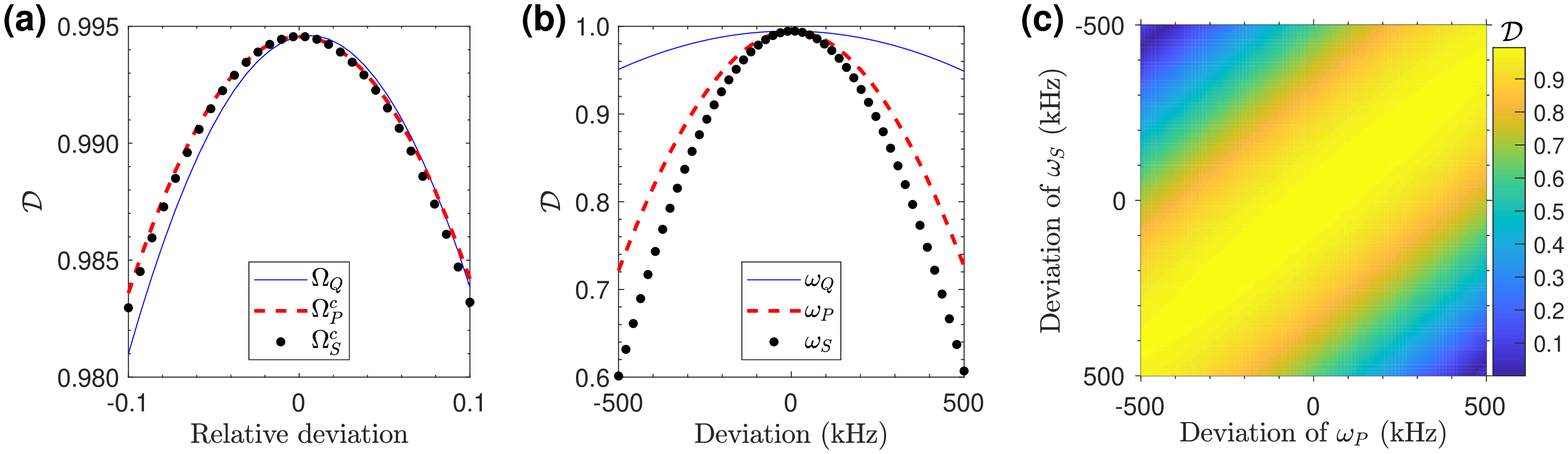}
\caption{(a)~Effect of the relative deviation of each pulse amplitude on the final enantio-discrimination. (b)~Effect of the deviation of each pulse frequency on the final enantio-discrimination. (c)~Joint effect of the frequency deviations of the $P$ and $S$ pulses on the final enantio-discrimination.
Each line in (a) and (b) involves only one parameter deviation. $\delta\omega_Q=0$ for (c). $dt=1$~ns and other parameters are the same as Fig.~\ref{f8}.}\label{f9}
\end{figure*}
We take into account the parameter deviations induced by frequency drifts and amplitude drifts that originate usually from the imprecise apparatus and imperfect operations, and define the deviation of an ideal parameter $x$ as $\delta x=X-x$ with $X$ being the actual value and the relative deviation as $\delta x/x$. Without losing generality, here we consider the relative deviation range $[-0.1,~0.1]$ of the pulse amplitudes and the deviation range $[-500,~500]$~kHz of the pulse frequencies. The effect of the relative deviation of each pulse amplitude on the final enantio-discrimination is plotted in Fig.~\ref{f9}(a), and the values of $\mathcal{D}$ keep always over $0.98$ within $\delta x/x\in[-0.1,~0.1]$ for all three pulses. The differences among three lines~(especially between $\Omega_P^c$ and $\Omega_S^c$) are quite slight. If the relative deviations of the pulse amplitudes can be restrained in $\delta x/x\in[-0.05,~0.05]$, the enantio-discrimination involving only one pulse amplitude deviation can be obtained finally with $\mathcal{D}>0.99$. So the present scheme is robust against pulse amplitude deviations.

The effect of the deviation of each pulse frequency on the final enantio-discrimination is plotted in Fig.~\ref{f9}(b). The pulse frequency deviations affects the final enantio-discrimination significantly, as the present scheme relies on the resonant regime strongly. And the frequency deviations should be restrained in $\delta x\in(-100,~100)$~kHz to enable $\mathcal{D}\gtrsim0.98$. The pulse frequency deviation of a pulse~(i.e., detuning of the corresponding transition) will spoil the resonant regime of the scheme. Usually, the damage extent of the resonant regime heightens with the increase of the ratio of detuning to Rabi frequency, but in Fig.~\ref{f9}(b) the line for the frequency deviation of $Q$ pulse with the lowest maximum amplitude holds the highest $\mathcal{D}$. The underlying physics locates at that $P$ and $Q$ pulses are related to not only the one-photon resonant processes but also the two-photon resonant process that determines the construction of 3MPPs. Even though the one-photon processes driven by $P$ and $Q$ pulses, respectively, are off-resonant, the high-$\mathcal{D}$ enantio-discrimination may also be achieved and the condition is $\delta\omega_P=-\delta\omega_S$, i.e., the two-photon resonant process, which is identified in Fig.~\ref{f9}(d). The $\delta\omega_P=-\delta\omega_S$ does not always ensure the high-$\mathcal{D}$ enantio-discrimination yet, and it works only when $|\delta\omega_P|$~($|\delta\omega_S|$) is not too large, about $|\delta\omega_P|<200$~kHz ensuring $\mathcal{D}>0.99$.

\subsubsection{Random amplitude noises}
\begin{figure}[htb]\centering
\includegraphics[width=0.9\linewidth]{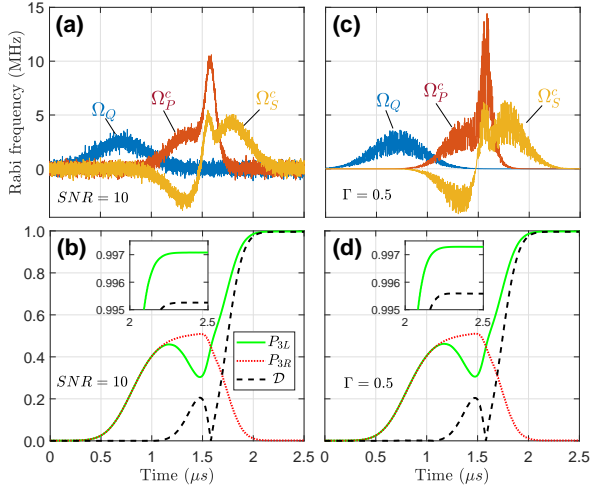}
\caption{(a)~Waveforms of three AWGN-mixed Rabi frequencies with $R_{\rm SN}=10$. (b)~Population evolutions of $|3\rangle$ of two enantiomers with $R_{\rm SN}=10$. (c)~Waveforms of three fluctuated-randomly Rabi frequencies with $\Gamma=0.5$. (d)~Population evolutions of $|3\rangle$ of two enantiomers with $\Gamma=0.5$. Parameters are the same as Fig.~\ref{f9}.}\label{f10}
\end{figure}
As for the pulse amplitudes, in addition to the systematic errors, there may exist noisy components in each pulse that disturb the intended dynamics. Different from systematic errors, these random noises are more unpredictable, and here we mainly consider two typical amplitude noises, AWGN and random fluctuations.
An AWGN-mixed Rabi frequency can be expressed by
\begin{eqnarray}\label{e23}
\Omega_{\rm AWGN}(t)&=&\Omega_{\rm ori}(t)+{\rm awgn}[\Omega_{\rm ori}(t),~R_{\rm SN}],
\end{eqnarray}
where ${\rm awgn}$ is a function that generates AWGN of the original pulse $\Omega_{\rm ori}(t)$~(i.e., $\Omega_Q$, $\Omega_P^c$ or $\Omega_S^c$) with a signal-to-noise ratio $R_{\rm SN}$.
A randomly-fluctuated Rabi frequency is written as
\begin{eqnarray}\label{e24}
\Omega_{\rm rand}(t)&=&\Omega_{\rm ori}(t)[1+{\rm rand}(t,~\Gamma)],
\end{eqnarray}
where ${\rm rand}$ denotes a function that generates a random number within $[-\Gamma,~\Gamma]$ at arbitrary time.
The waveforms of three AWGN-mixed Rabi frequencies with $R_{\rm SN}=10$ are shown in Fig.~\ref{f10}(a), based on which the population evolutions of $|3\rangle$ of two enantiomers are plotted in Fig.~\ref{f10}(b). We can clearly see from Fig.~\ref{f10}(b) that the enantio-discrimination is influenced little by the AWGN with the signal-to-noise ratio $R_{\rm SN}=10$. When the random fluctuations of the pulse amplitudes are taken into account, even with $\Gamma=0.5$, the waveforms are shown in Fig.~\ref{f10}(c), and a high-$\mathcal{D}$~(0.9956) enantio-discrimination is still reached, as shown in Fig.~\ref{f10}(d). Generally, the scale of noise is much smaller than the scale of the original pulses. These results indicate that the influences of the random amplitude noises including AWGN and random fluctuations can be neglected for the enantio-discrimination, for which the reason is due to the fact that the time average effect of these random noises is zero. AWGN and random fluctuations possess random absolute values and random plus-minus signs, so the collective effect on the pulse areas is little and the intended dynamics is distortionless.

\subsection{Energy relaxation}
\begin{figure}[htb]\centering
\includegraphics[width=0.9\linewidth]{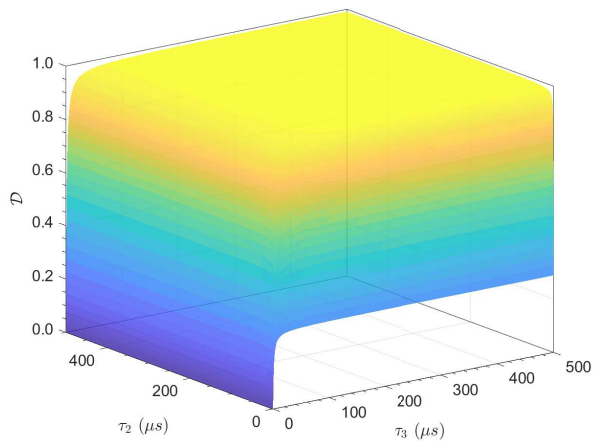}
\caption{Final enantio-discrimination with varying $\tau_2$ and $\tau_3$. Parameters are the same as Fig.~\ref{f9}.}\label{f11}
\end{figure}
For convenience of the discussion above, we neglect the energy relaxations of higher states by setting $\gamma_{1,2}=\gamma_{1,3}=\gamma_{2,3}=0$. Although the considered closed-loop configuration consists of the relatively lower rotational energy states that have relatively longer lifetime, there is the probability yet that two higher states $|2\rangle$ and $|3\rangle$ may relax to the states lower than them. We suppose the lifetimes of $|2\rangle$ and $|3\rangle$ as $\tau_2$ and $\tau_3$, respectively, and then $\gamma_{1,2}=1/\tau_2$ and $\gamma_{1,3}=\gamma_{2,3}=0.5/\tau_3$ for simplicity.
Then with varying $\tau_2$ and $\tau_3$, the final enantio-discrimination is depicted in Fig.~\ref{f11}. Figure~\ref{f11} exhibits that the shorter the lifetime of $\tau_2$ or $\tau_3$ is, the more significantly the final enantio-discrimination is spoiled, but the near-full~($\mathcal{D}>0.99$) enantio-discrimination can be always acquired as long as $\tau_2>200~\mu s$ and $\tau_3>300~\mu s$ which are accessible for the states $|2\rangle\equiv|1_{11}\rangle$ and $|3\rangle\equiv|1_{10}\rangle$ in cold $1,2$-propanediol molecules. Therefore, the effect of decoherence induced by energy relaxations on the present enantio-discrimination scheme is insignificant.

\section{Conclusion}\label{S6}
Based on the combination of a one-photon process and a two-photon process in a closed-loop configuration, we have proposed to realize the robust and highly-efficient discrimination of chiral molecules. The one-photon process is responsible for the coherence between $|1\rangle$ and $|3\rangle$, and then the two-photon process through three-mode parallel paths constructed by the method of ``chosen paths" enables two enantiomers evolving to entirely-different states, i.e., a full enantio-discrimination. By means of the pulse engineering, many kinds of pulse shapes can used in implementing the scheme. The one- and two-photon processes can overlap partly with each other, which shows the robustness against unfaithful  pulse delays and can further shorten operation time. The phase sensitivity of the scheme indicates the enantio-discrimination can be obtained by setting the phase of any one pulse.
Besides, we propose to apply the present scheme in $1,2$-propanediol molecules, and the experimental issues are considered including unwanted transitions, imperfect initial state, pulse shaping, control errors and the effect of energy relaxations. With the unwanted transitions and imperfect initial state, in a closed-loop configuration involving three lower rotational states of $1,2$-propanediol molecules, the robustness of the scheme against the long time resolution, amplitude drifts, frequency drifts and random noises of pulses is illustrated. The scheme is also hardly influenced by energy relaxations since the used lower rotational states are of long coherence time.

Further, the present scheme may be optimized by means of the state-of-the-art quantum optimal control techniques~\cite{PhysRevLett.104.083001,Glaser2015,PhysRevA.85.023611,PhysRevLett.111.050404,PhysRevA.88.043422,PhysRevA.95.063403,PhysRevA.98.043421} which can give optimal plans for one or more particular purposes, and the combination~\cite{Mortensen_2018,PhysRevA.99.022327,PhysRevA.99.063812,PhysRevLett.123.100501} of quantum optimal control with STA or STIRAP may provide more excellent results in efficiency, robustness, and accuracy of carrying out enantio-discrimination tasks.
Finally, we hope that our work could provide the substantial help for laboratory researchers in the separation of chiral molecules.

\section*{ACKNOWLEDGEMENTS}
All authors would like to thank the anonymous referees for constructive comments that are helpful for improving the quality of the work. This work was supported by National Natural Science Foundation of China (NSFC) (11675046), Program for Innovation Research of Science in Harbin Institute of Technology (A201412), and Postdoctoral Scientific Research Developmental Fund of Heilongjiang Province (LBH-Q15060).

\bibliography{apssamp}
\end{document}